# Not Dead, Just Resting: The Practical Value of Per Publication Citation Indicators[1]

Mike Thelwall, University of Wolverhampton.

## 1. Introduction

Citation counts do not *measure* research impact under any reasonable definition of impact. Instead they are *indicators* of research impact or quality in the sense that in some disciplines, when appropriately processed, they tend to correlate positively and statistically significantly with human judgements of impact or quality. Although we have theory to suggest that, at least in the sciences, citations could tend to reflect the contributions of papers to the onward progress of scholarship (Merton, 1973), there are many examples to demonstrate that this is not always the case (MacRoberts & MacRoberts, 1989, 1996, 2010; Seglen, 1997). Moreover, all citation-based indicators are often wrong and misleading and for some fields are completely useless (e.g., music and art in Table A3 of: HEFCE, 2015a). Nevertheless, as argued below, as long as they have a significant positive correlation with human judgements then they can serve a useful role supporting peer review (see also: van Raan, 1998). They may even be a primary source of evidence in situations where the expense of peer review in comparison to its benefits makes it impractical or for cases where peer judgements are thought to be too biased. Thus, citation-based indicators should not be used as a primary arbitrator of research impact unless there are practical reasons why better alternatives are inappropriate. Peer review is, in general, a better solution and is the one used in the UK to direct a large amount of UK government research funding (£1.6 billion per year in 2014-15: Wilsdon, Allen, Belfiore, Campbell, et al., 2015) with the remainder being primarily distributed through competitive project grant applications. The Abramo and D'Angelo (2016) argument for a particular type of indicator should be analysed within this context: not as an attempt to construct a perfect measure of research impact but as an attempt to construct the best possible indicator of research impact that is based on citations.

When counting citations to collections of papers in order to compare two or more heterogeneous sets of papers, it is important to normalise the counts in order to make the comparisons fairer. For example, medical research is cited more frequently than information science research and older papers have had longer to attract citations. One way of reducing biases is to divide each article by the average number of citations for other articles of the same type and publication year so that scores above 1 suggest that a publication has had above average impact, irrespective of its field and publication year. The Mean Normalised Citation Score (MNCS) (Waltman, van Eck, van Leeuwen, Visser, & van Raan, 2011a,b) extends this to sets of documents by calculating the arithmetic mean of the normalised citation score for each publication so that an MNCS score above 1 indicates that the collection of papers has had an above average citation impact, irrespective of the range of publication years, fields and document types. There have been suggestions for fine tuning aspects of this calculation, such as by replacing the arithmetic mean with the geometric mean for its better handling of highly skewed data (Fairclough & Thelwall, 2015), or to report instead the proportion of articles in the most highly cited percentile (e.g., top 10% or top 1%) (Tijssen, Visser, & Van Leeuwen, 2002) but neither of these affect the main discussion here. This is because the main issue is whether indicators like these that are calculated on a per publication basis can be useful rather than which of them is the best.

The MNCS and related indicators are used in applied scientometrics as evidence to aid peer review and to support policy evaluations and decisions. In this context, Abramo and D'Angelo (2016) argue that they are "not worthy of further use or attention". The root problem is that averaging citation counts across exhaustive sets of articles to be compared can lead to misleading results, as they clearly demonstrate. For example, if there are ten sets of articles, each of which contains all of the publications of a given Dutch physics research group 2011-2015 and the purpose is to decide how much government funding should be given to each group during 2016-2020 then groups would be

---

penalised for producing below average impact articles because these lower their average score. One group might have a higher total number of citations but a much lower MNCS because it produced low impact articles in addition to high impact articles, whereas another group produced the same amount of high impact articles but fewer low impact articles. In this case the group with the most total citations and the same number of high impact papers would have the lowest MNCS, unfairly penalising it for doing more work (Abramo & D'Angelo, 2016). The underlying problem here is that the size of the groups producing the articles is ignored in the MNCS calculation. This problem could be reduced if a degree of filtering was allowed in the selection of articles to be averaged. Researchers in the UK are evaluated on the basis of their best four outputs over a period of about six years (Wilsdon, et al., 2015). Averaging the citation impacts of a set of articles collected on this basis would reduce the problem of one group's "additional" low-cited articles reducing its average citation count. Other problems would remain, however, because groups producing more than four articles per researcher would not get rewarded for this (which is accepted in the UK and has probably increased quality at the expense of quantity: Moed, 2008).

In the above Dutch physics groups example the indicators are clearly misleading and this is not just a theoretical problem but also occurs in practice (Abramo & D'Angelo, 2016). Abramo and D'Angelo (2016) argue that the fundamental issue underlying the misleading scores in the example above is that citation indicators should not be independent of the resources used to generate the research but should normalise by a measure of research group size or input, such as researcher numbers or money. In other words, comparing the field normalised average number of citations per researcher or the field normalised average number of citations per Euro of funding would be preferable to the field normalised average number of citations per paper. The same logic holds for other indicators based on the ratio to the total number of publications, such as the percentage of a set of articles in the most cited X% for a field and year. I agree with the logic of their argument but disagree with their conclusion that ratio to publication indicators should never be used. I also think that there are practical problems that will prevent indicators like MNCS being replaced by funding or personnel normalised indicators for the foreseeable future in many, and probably most, contexts. These problems are the incompatibility of personnel or finance data with the functioning of the systems being compared and the additional expense of gathering personnel or finance data outweighing the additional value provided. These are described below for a few areas in which the MNCS or variants are currently used.

## 2. Per publication indicators are not useless

All citation-based indicators have problems (MacRoberts & MacRoberts, 1989, 1996, 2010; Seglen, 1997) and so demonstrating the problems of per publication indicators is not enough to claim that they give information of "no value to decision-makers" (Abramo & D'Angelo, 2016). As long as any indicator's values tend to correlate positively and statistically significantly with human judgements it is giving *some* information (Moed, 2008; see also: van Raan, 1998). For example, if a group of physics experts are given MNCS scores for the Dutch physics groups and are told the limitations of the indicator then this tells them the broadly expected average impact of the research conducted by the groups at a level above chance. The experts can then use the MNCS values to cross-check their judgements, ensuring that they have not overlooked emerging areas of excellence or overestimated the value of a particular group. In this context, MNCS data could *warn* the experts to take a second look at groups with anomalous MNCS results, whilst allowing their own judgements to be the final arbiter.

Indicators, however flawed, that correlate positively with human judgements can also be used for systematic cross-checking of evaluation results or the evaluation process. This can be used to look for evidence of bias within the judgments, such as in terms of gender, ethnicity or seniority (e.g., Ceci & Williams, 2011; HEFCE, 2015ab; Østby, Strand, Nordås, & Gleditsch, 2013). Of course, the results of such analyses must be followed up to investigate whether any positive scores are due to biases in the indicator itself.

Accepting that per publication indicators are not useless, they should still be ignored if they can be replaced by per researcher or per Euro indicators that are clearly better. Counting researchers or funding can be difficult and this could be seen as a challenge to scientometricians (Abramo & D'Angelo, 2016). Nevertheless, the value of the extra information given by the more accurate indicators can sometimes outweigh the cost of calculating them and per researcher and per Euro indicators may be more easily gamed. Specific cases are discussed below.

## 3. Departmental funding-related evaluations may not be able to calculate per Euro or per researcher indicators that cannot be gamed

When evaluating a set of departments or other research units from the same field, the MNCS values may be calculated for each department and given to the expert evaluators in order to aid their decision making. The purpose of such an exercise may be to allocate future block grants or to decide which departments to close. Normalisation by the finance given to a research unit seems to be impractical in many countries, such as the UK. This is because much of the finance given to a research unit is an output of its activities rather than an input. A research group *wins* research grants after extensive work on writing applications and forming consortia, and the winning of research grants is seen as evidence of research excellence. For normalisation purposes it would be possible to consider only the amount of block grant previously allocated (although this may also have been won in a previous round) and ignore all funding from project grants. This is also problematic because some universities (e.g., Harvard, Cambridge) fund research partly through their historically-generated resources, which is also unfair and would have a substantial impact on the results and would be demoralising for departments in newer, poorly-financed institutions. New research groups formed with university money rather than from government block grants would also cause a problem because of their denominator of zero (1 citation and no government funding = infinite impact per Euro).

Normalisation by the number of researchers is also problematic because it is very difficult to count the number of active researchers in a fair way even within a single country. In the UK, people with the job title "Lecturer" may have contracts that allow them to spend 50% of their time on research or give them only a small allowance for "scholarly activity" but no time for research. Universities may also award individually negotiated time off from teaching for research. Hence, in the UK there are probably lecturers that are contracted to conduct any amount of research from 0% to 100%. Thus, calculating the fractional total number of person-years devoted to research in a department would be a highly complex undertaking. Moreover, if this statistic would become important enough to be used in evaluations then it would be trivial to game contracts so that all "research" became "scholarly activity" and denominators would shrink dramatically. This considers only one grade of academic, the lecturer, and does not consider the many other types: postdocs, researchers, demonstrators, professors, assistant professors, readers, and associate professors as well as support staff that have some input into research, such as lab technicians. Research is so enormously varied and intertwined with teaching in many universities that it seems a highly complex task to count researchers and one that would be impossible if there was stakeholder interest in getting a low answer for a field normalised citation indicator.

In this case the problems are very difficult to solve in a reasonable way but probably not impossible in the long run, such as through standardisation. Nevertheless, the problems seem to be substantial enough that it seems likely that the cost and potential harm to university research needed to achieve standardisation would far outweigh the benefits of a slightly more rigorous indicator in the UK (where citations play a very minor role in departmental evaluations, and are only used in a minority of subjects: Wilsdon, et al., 2015) and all countries except any that already have a homogenous university system with strictly defined roles without a variable research component. Alternatively, if citation indicators are needed for purposes that do not have major stakeholder interests in a particular result, such as self-evaluations, that include sets of comparators then existing block grant data and information about numbers of researchers might be acceptable.

## 4. University ranking systems may not be able to use per Euro or per researcher indicators that cannot be gamed

The considerations above for departments also apply to comparisons between universities, if these are ever used for funding evaluations. Similarly, although international university rankings (Aguillo, Bar-

Ilan, Levene, & Ortega, 2010) tend not to drive funding directly they are important for prestige, which can help to attract students, non-academic funding, and higher quality researchers (e.g., Wilsdon et al., 2015). Hence, universities can be expected to make adjustments to improve their scores in the well-known ranking websites, whatever their flaws (Salmi, 2015). This is almost certainly happening already on a large scale but highly-weighted per Euro or per researcher indicators in any ranking system would add to the problem.

For international university rankings, counts of income and research personnel are unlikely to be internationally comparable in the foreseeable future, as argued in the section below.

# 5. International per Euro or per researcher indicators may not be comparable

Governments sometimes compare their research base with the research bases of other countries in order to help assess national progress or recent policy changes (NIFU, 2014; NSF, 2014; Salmi, 2015). For this, they may use a range of sources of evidence, one of which being the citations to the articles produced by the country and a set of comparators using a variant of the MNCS (e.g., Elsevier, 2013). Although it would be possible to compare on the basis of citations per dollar, and this might give a useful additional indicator to MNCS-like values, this would not clearly be better because, except for the Eurozone, it would depend on exchange rates and the international economic strengths of the countries concerned and so it is not clear how useful the information would be. Another problem is that some countries have privately-funded universities as a core part of their university system (e.g., the USA) whereas in others, higher education is almost exclusively publically-owned (e.g., the UK).

If countries are compared on the basis of citations per researcher then this would only be reasonable for sets of countries with comparable systems. For example, Germany has many research-only institutes (Max Planck Society, Helmholtz Association) and Spain has the large national research council (CSIC). Including these researchers without substantial educational commitments seems to give an unfair advantage over countries that do not have equivalent organisations. On the other hand they are part of the overall research ecosystem of the country and so leaving them out also seems to be unfair.

The nature of organisations that are accepted as being higher education also changes over time, affecting the number of people that may be classified as researchers if a broad definition is used. After September 2013 in the UK nursing qualifications changed from diploma level to degree level and other professions have also experienced a transition from non-degree to degree-level qualifications. More generally, most countries seem to be expanding their higher education systems, changing the nature of those who work in them. In addition, some countries have organisations called universities that are teaching intensive (e.g., the USA, the UK) whereas others have vocational higher education institutions that may not be called universities (e.g., Dutch *hogescholen* are technical high schools but may also be called universities of applied sciences) and the line between higher education and further education may be blurred (e.g., further education colleges in the UK may offer external degrees or foundation courses for degrees).

In the context of the international complexity of higher education and research systems, it would be difficult to claim any degree of robustness in the numbers of researchers employed by any particular country. Given that governments may not be willing to spend much on citation-based indicators because they are only one of a number of sources of information used (e.g., Elsevier, 2013; NIFU, 2014; NSF, 2014; Salmi, 2015), and the countries that they wish to be comparators may have very different education systems (e.g., richer countries might like to see China, the USA and Japan as comparators) it seems unlikely that effective citations-per researcher indicators could be generated. Nevertheless, it would be a good idea to produce them in addition to the MNCS when comparators have similar systems and a convenient source of data on researcher numbers is available.

An alternative approach is to calculate country figures on a national scale per inhabitant (Salmi, 2015), which may give broadly reasonable figures that have the advantage of being transparent and perhaps reasonable for similar types of country. Funding can also be taken into account by comparing the per-inhabitant performance of countries with the percentage of Gross Domestic Product (GDP) spent on research (Salmi, 2015).

## 6. Summary

In the final analysis citation-based indicators are inferior to effective peer review and even peer review is flawed. It is impossible to accurately measure the value or impact of scientific research and a key task of scientometricians should be to produce figures for policy makers and others that are as informative as it is practical to make them and to ensure that users are fully aware of their limitations. Although the Abramo and D'Angelo (2016) suggestions make a lot of theoretical sense and so are a goal that is worth aiming for, it is unrealistic in practice to advocate their universal use in the contexts discussed above. This is because the indicators would still have flaws in addition to the generic limitations of citation-based indicators and would still be inadequate for replacing peer review. Thus, the expense of the data gathering does not always justify the value in practice of the extra accuracy. In the longer term, the restructuring of education needed in order to get the homogeneity necessary for genuinely comparable statistics would be too expensive and probably damaging to the research mission, in addition to being out of proportion to the likely value of any citation-based indicator.